\documentclass[aps,prl,twocolumn,groupedaddress,showpacs]{revtex4}
\usepackage{graphicx, amsmath,bm}

\newcommand{\bs}[1]{{\boldsymbol{#1}}}   

\begin{document}
\title{A topological look at the quantum spin Hall state}
\author{Huichao Li}
\author{L. Sheng}
\email{shengli@nju.edu.cn}
\author{D. Y. Xing}
\affiliation{ National Laboratory of Solid State Microstructures and
Department of Physics, Nanjing University, Nanjing 210093, China}

\begin{abstract}
We propose a topological understanding of the quantum spin Hall
state without considering any symmetries, 
and it follows from the gauge invariance that either the
energy gap or the spin spectrum gap needs to close 
on the system edges, the former scenario generally resulting
in counterpropagating gapless edge states. Based upon the Kane-Mele model
with a uniform exchange field and a  
sublattice staggered confining potential
near the sample boundaries, we demonstrate the existence
of such gapless edge
states and their robust properties in the presence of impurities.
These gapless edge states
are protected by the band topology alone, rather than any symmetries.

\end{abstract}
\pacs{72.25.-b, 73.20.At, 73.22.-f, 73.43.-f}
\maketitle
Since the remarkable discovery of the quantum Hall effect
(QHE)~\cite{QHE},  the study of edge state physics has attracted
much attention on both theoretical and experimental sides. Recently,
a new class of topological states of matter has emerged, called the
quantum spin Hall (QSH) states~\cite{Kane1,Bernevig}. A QSH state of
matter has a bulk energy gap separating the valence and conduction
bands and a pair of spin-filtered gapless edge states on the
boundary. The QSH effect was first predicted in two-dimensional (2D)
models~\cite{Kane1,Bernevig}, and was experimentally confirmed soon
after in mercury telluride quantum wells ~\cite{Konig}. The QSH
systems are 2D topological insulators~\cite{TI1,TI2} protected by the time-reversal
symmetry (TRS), whose edge states are robust against perturbations
such as nonmagnetic disorder.

A simple model of the QSH systems is the Kane-Mele
model~\cite{Kane1}, defined on a honeycomb lattice,
first introduced for graphene with spin-orbit
couplings (SOCs). It was suggested~\cite{Kane1} that the intrinsic SOC in graphene
would open a band gap in the bulk and also
establish spin-filtered edge states that traverse the band gap,
giving rise to the QSH effect. Even though the intrinsic SOC
strength in pure graphene is too small to produce an
observable effect under realistic conditions~\cite{Min}, the
Kane-Mele model captures the essential physics of the QSH state
with nontrivial band topology~\cite{Rev1,Rev2}. 
In the presence of the Rashba SOC and
an exchange field, the Kane-Mele model enters a TRS-broken QSH
phase~\cite{sheng} characterized by nonzero spin Chern
numbers~\cite{spinchern,Prodan}. Prodan proved~\cite{Prodan} that
the spin-Chern numbers
are topological invariants, as long as the energy gap and the spectrum gap of the 
projected spin operator $P\sigma_zP$
stay open in the bulk, where $P$ is the projection operator onto the subspace of
the occupied bands and $\sigma_z$ the Pauli matrix
for the electron spin. Unlike the $Z_2$
invariant~\cite{z2}, the robust properties of the spin-Chern numbers
remain unchanged when the TRS is broken~\cite{sheng,Prodan}.

The existence of counterpropagating edge states with opposite spin
polarizations is an important characteristic of the QSH state. 
It is believed that the edge states can be gapless only
if the TRS~\cite{Kane1} or
other symmetries, such as the inversion symmetry~\cite{BZhou}
or charge-conjugation TRS~\cite{QFSun}, are present. When the TRS
is broken, it was found~\cite{sheng} that a small gap
appears in the spectrum of the edge
states, which was obtained for a ribbon geometry under ideal
boundaries, i.e., boundaries created by an infinite hard-wall
confining potential. However, since the edge states are localized around
the sample boundaries, they can be sensitive to the variation of
on-site potentials near the boundaries~\cite{niu1}.

In this Letter, in order to 
reveal the general characteristics of the edge states
and their connection to the bulk topological invariant in a QSH
system, we present a topological argument similar to the
Laughlin's Gedanken experiment without considering any symmetries. 
We show that, as
required by the nontrivial band topology and gauge invariance, 
either the energy gap or the spin spectrum gap (the gap in the spectrum of
$P\sigma_zP$) needs to close on the
edges of a QSH system. These two scenarios will lead to gapless or
gapped edge modes, respectively. In particular, it is demonstrated
that gapless edge states can appear in a TRS-broken QSH system by
tuning the confining potential at the boundaries. They
are associated with the bulk topological invariant, and
are robust against relatively smooth impurity scattering potential.
Our result offers an interesting example for 
counterpropagating gapless edge states that are
not protected by symmetries, which sheds light on
the underlying mechanism of the QSH effect in a broad sense.

Let us first look back on a looped ribbon of the QHE system, with a
magnetic flux $\phi$ (in units of flux quantum $hc/e$)
threading the ring adiabatically~\cite{new1,new2,new3}. The Fermi
energy $E_F$ is assumed to lie in an energy gap. In the
spirit of the Laughlin's argument, increasing $\phi$ from 0 to
$1$ effectively pumps one occupied state from one edge to the
other, giving rise to the transfer of one charge between the edges,
essentially because there is a nonzero Chern number (Hall
conductivity) in the bulk. On the other hand, the system Hamiltonian
is gauge invariant under integer flux changes, i.e., if $\phi$ is
increased from $0$ to $1$, the system will reproduce the same
eigenstates as at $\phi=0$. To assure this gauge invariance, there
must be gapless edge modes on the edges (when the perimeter of the
ring is large), so that the spectral flow can form a closed loop, as
illustrated in Fig.\ 1a, along which the electron states can
continuously move with changing $\phi$.

\begin{figure}[htbp]
\includegraphics[width=2.7in]{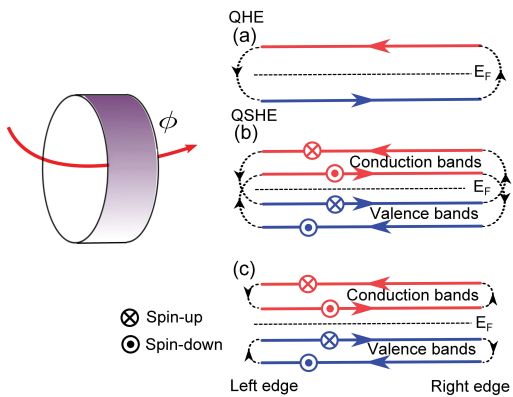}
\caption{(a) A schematic of the flow of electron states in a 
looped ribbon of the QHE
system, with adiabatically increasing the magnetic flux
$\phi$ that threads the ring. The bulk electron states below Fermi
energy $E_F$ drift from left to right, gapless edge modes ascend
through $E_F$ on the right edge, states above $E_F$ drift from right
to left, and gapless edge modes descend through $E_F$ on the left
edge, forming a closed loop. (b) In the first scenario for the QSH
system, gapless edge modes appear on the edges of the ring, so that
electron states in each spin sector behave like in the QHE system,
but those in two spin sectors move in opposite directions. (c) In
the second scenario  for the QSH system, the edge states are gapped,
whereas the spin spectrum gap closes on the edges. In the bulk,
electron states in the two spin sectors drift in the same way as in
(b), but on the edges they join together within the valence
(conduction) band. } \label{Laughlin}
\end{figure}
We now propose a topological understanding of the QSH system in terms of
the same looped ribbon geometry. The occupied valence band can be decomposed
into two spin sectors by using the projected spin operator~\cite{Prodan,sheng}. 
(The unoccupied conduction band can be divided similarly.) The
two spin sectors are separated by a nonzero spin spectrum gap
in the bulk. They 
carry opposite spin Chern numbers, so that increasing $\phi$ pumps
a state of the spin up sector in the occupied band 
from one edge to the other, and 
pumps another state of the spin down sector in the opposite
direction. In order for the system to recover the initial
eigenstates as $\phi$ changes from $0$ to $1$, the spectral flow
needs to form closed loops, similarly to the QHE system. 
However, for the QSH system, if not
enforced by any symmetry, two different scenarios 
can occur on the edges.
One is that gapless edge modes appear on the edges, so that states
can move between the conduction and valence bands with changing
$\phi$ to form closed loops in the spin-up and spin-down sectors
separately, as shown in Fig.\
1b. In this case, the states in the two loops cannot evolve into each
other due to the nonvanishing spin spectrum gap both in the bulk and 
on the edges. The other scenario
is that a closed loop of spectral flow is formed between the two spin
sectors within the valence (or conduction) band, as shown in Fig.\ 1c. 
In this case, the spin spectrum
gap must vanish on the edges, but the energy gap may remain open
in the edge state spectrum. 

\begin{figure}[htp]
\includegraphics[width=2.7in]{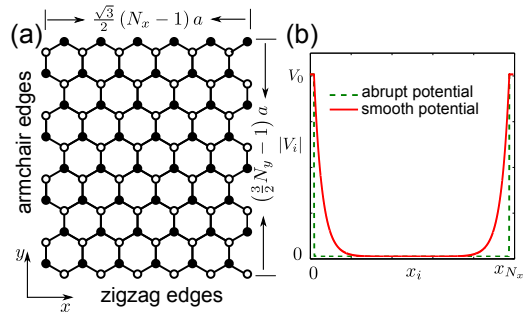}
\caption{(a) A schematic of an armchair honeycomb lattice
ribbon with atom sites in two sublattices being labeled by solid
dots and hollow dots, where $a$ is the distance between
nearest-neighbor sites. (b) Profiles of $\vert V_i\vert$ as functions of
$x_i$, for an abrupt confining potential (dashed line) and a relatively 
smooth confining potential (solid line).} \label{arm}
\end{figure}
The above topological discussion on the QSH system is very general,
independent of any symmetries. To demonstrate the two
scenarios in Figs.\ 1b and 1c, in what follows we take the Kane-Mele
model~\cite{Kane1} for a honeycomb lattice ribbon as an example, by 
taking into account 
different confining potentials near the edges of the 
ribbon. It was shown that in a suitable parameter range, the
Kane-Mele system is in the QSH phase protected by the TRS, and it
can become a TRS-broken QSH phase~\cite{sheng}, when a spin-splitting
exchange field is applied. Consider an armchair ribbon along
the $y$ direction, as shown in Fig.\ 2a, including $N_x$ dimer lines
across the ribbon. (Results for a zigzag ribbon are similar.)
The boundaries are at $x_1 = 0$ and
$x_{N_x}=\frac{{\sqrt 3}}{2}(N_x-1)$, where the distance between 
nearest-neighbor
sites is chosen as the unit of length. 
The Hamiltonian can be written as $H=H_{KM} + H_{E} +H_{C}$ with
\begin{eqnarray}\label{ham}
H_{KM} &=&  - t\sum\limits_{ \langle ij \rangle} {c_i^\dag {c_j}} +
\frac{2i}{{\sqrt 3 }}V_{SO}\sum\limits_{\langle\langle ij
\rangle\rangle}
{c_i^\dag \bs{\sigma} \cdot({\bs{d}_{kj}} \times {\bs{d}_{ik}}){c_j}}  \nonumber\\
&+& i{V_{R}}\sum\limits_{\langle ij \rangle} {c_i^\dag
{{\hat{\bs{e}}}}_z\cdot(\bs{\sigma}\times {\bs{d}_{ij}}){c_j}},
\end{eqnarray}
as the Hamiltonian of the Kane-Mele model. Here, the first term is
the nearest-neighbor hopping term with $c_i^\dag = (c_{i \uparrow
}^\dag ,c_{i \downarrow }^\dag )$ as the electron creation operator
on site $i$ and the angular bracket in $\langle i,j\rangle$ standing
for nearest-neighbor sites. 
The second term is the intrinsic
SOC with coupling strength $V_{SO}$, where
$\mbox{\boldmath{$\sigma$}}$ are the Pauli matrices, $i$ and $j$ are two next
nearest neighbor sites, $k$ is their unique common nearest neighbor,
and vector ${\bf d}_{ik}$ points from $k$ to $i$. The third term is
the Rashba SOC with coupling strength $V_{R}$. $ H_{E} =g\sum_i
{c_i^\dag {\sigma _z}{c_i}}$ stands for a uniform exchange field of
strength $g$.  $H_{C}$ represents a sublattice staggered 
confining potential, which is given by 
$H_{C}= \sum_i {{V_i}c_i^\dag{c_i}}$ with
\begin{equation}
V_i = \pm {V_0}\left( {{e^{ - \frac{x_i}{\xi }}} + {e^{ -
\frac{{x_{N_x}- x_i}}{\xi }}}} \right)\ ,
\end{equation}
where $\pm$ is taken to be positive for sites
on sublattice $A$ (solid dots) and negative on
sublattice $B$ (hollow dots), as shown in the Fig.\ \ref{arm}a. 
In Eq.\ (2), $V_i$ is strongly
$x$ dependent across the ribbon, equal to $\pm V_0$ at the
edges ($x_i=0$ and $x_i=x_{N_x}$). It decays exponentially
away from the edges, with a
characteristic length $\xi$, as
shown in Fig.\ \ref{arm}b. When the ribbon width is much greater
than $\xi$, $V_i$ essentially vanishes in the middle region of the ribbon. 
Here, we note that in the
case of a uniform staggered potential $V_i=\pm V_0$ ($\vert V_i\vert$ being 
independent of $x_i$), it was shown~\cite{sheng} that with increasing 
$\vert V_i\vert$, there is a transition from the TRS-broken QSH phase to an
ordinary insulator state, where the middle band gap closes and then reopens. 
Therefore, for the
confining potential $V_i$ given by Eq.\ (2) with large $V_0$, the ribbon in Fig.\
\ref{arm} can be regarded as a TRS-broken QSH ribbon sandwiched in
between two trivial band insulators.


\begin{figure}[htp]
\includegraphics[width=3.2in]{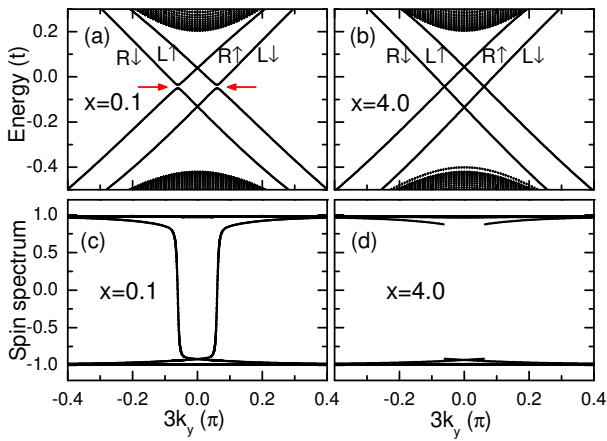}
\caption{(color online) The energy spectrum (a, b) and the spectrum
of the projected spin $P\sigma_zP$ (c, d) of an
armchair ribbon for $\xi =0.1$ (a, c) and $\xi=4$ (b, d),
in which $L$ ($R$) stands for states on the left (right) edge,
and $\uparrow$ ($\downarrow$) for the up (down) spin polarization. 
The horizontal arrows in (a) point to the small energy gaps in
the energy spectrum. 
At $\xi=0.1$, while the energy spectrum (a) is gapped, 
the spin spectrum (c) is gapless. At $\xi=4$,
the energy spectrum (b) is gapless, but the spin spectrum (d) is gapped.
} \label{edge1}
\end{figure}

In order to assure the system in the TRS-broken QSH state, we set
the parameters $V_{so}=0.1t$, $V_R=0.1t$, and $g=0.1t$. 
The length of the armchair ribbon $N_y$ is taken to be infinite. 
The energy spectrum of the
ribbon, together with the corresponding
eigenfunctions $\varphi_m(k_y)$, can be numerically obtained by
diagonalizing the Hamiltonian for each momentum $k_y$ 
in the $y$ direction. The calculated energy spectrum of the armchair
ribbon with width $N_x=240$, for the confining potential with 
$V_0=12t$ fixed and two different decay lengths,
is plotted in Fig.\ \ref{edge1}a and Fig.\
\ref{edge1}b. One can see easily that
 the edge states appear as thin lines in
 the middle bulk band gap of the energy spectrum. In Figs.\ \ref{edge1}a and
\ref{edge1}b, the spin polarization of the edge states is labeled
with $\uparrow$ and $\downarrow$, 
indicating that two spin-filtered channels
on each edge flow along opposite directions. For the nearly hard-wall
confining potential of $\xi=0.1$, 
the sublattice
potential $V_i$ is nonzero only on the outermost armchair lines, 
similar to
the assumption in Ref.\ \cite{niu1}. In this case, 
two small energy gaps are observed in the edge 
state spectrum  shown in Fig.\ \ref{edge1}a, in agreement with 
the previous observation~\cite{sheng}, as a consequence
of the broken TRS. With increasing the decay length $\xi$ of
the confining potential, the energy gaps
of the edge states become smaller and smaller. As $\xi$ is large
enough, e.g., $\xi=4$ in Fig.\ \ref{edge1}b, interestingly, the
edge states become gapless.

We now calculate the $k_y$-dependent spectrum of 
projected spin operator $P\sigma_zP$, whose matrix
elements are given by 
${\left\langle {{\varphi
_m\left( k_y \right)} }\right|{\sigma_z}\left| {{\varphi_n}\left(
k_y \right)} \right\rangle}$
with $m$ and $n$ running over all the
occupied states. By diagonalizing this matrix, the spectrum
of the projected spin $P\sigma_zP$ can be obtained. 
For the Kane-Mele model Eq.\ (1), if
$V_R=0$, $\sigma_z$ commutes with the Hamiltonian $H$. Therefore,
$\sigma_z$ is a good quantum number. It follows that
the spectrum of $P\sigma_zP$ consists
of just two values $\pm 1$, which are highly degenerate. When the
Rashba term is turned on, $\sigma_z$ and $H$ no longer commute, and the
degeneracy is lifted. In this case, the spectra of $P{\sigma_z}P$ between
$+1$ and $-1$ spread towards the origin, but a gap remains for a bulk sample 
if the amplitude of the Rashba term dose not exceed a
threshold~\cite{Prodan}. 

For the ribbon geometry, the situation is more
complicated due to the existence of the edges, and numerical
calculations are performed to obtain
the spin spectrum. The calculated spectrum of $P{\sigma_z}P$
for the same parameters as those in Figs.\ 3a and
3b is shown in Figs.\ 3c and 3d, which exhibits very
interesting behavior. For the hard-wall confining
potential with $\xi=0.1$, while the energy spectrum of edge states are
slightly gapped, with increasing $k_y$ the spectrum of $P\sigma_zP$ 
continuously change between $+1$ to $-1$ without showing any gap, 
corresponding to the second scenario
shown in Fig.\ 1c. On the other hand, for a relatively smooth confining
potential with $\xi=4.0$, the energy spectrum of the edge states are
gapless, but the spectrum of $P\sigma_zP$ displays a big gap, and the
sudden changes happen at the cross points in the energy spectrum
of the edge states, corresponding to the
first scenario shown in Fig.\ 1b. From Fig.\ 3, it follows that as
long as the system is in the QSH state, a gapless characteristic
always appears either in the energy spectrum of edge states or in
the spectrum of $P\sigma_zP$, leading to the two types of closed loops for
the continuous flow of the electron states illustrated in
Figs.\ (1b, 1c).

The result shown in Figs.\ 3b and 3d, for the relatively 
smooth confining potential, 
is of particular interest. It indicates that gapless edge states
can exist in the TRS-broken QSH system,
accompanied with a gapped spectrum of $P\sigma_zP$. Such an
interesting behavior can be further understood by the following argument.
As long as the bulk energy gap does not close, the projected spin
operator $P\sigma_zP$ is exponentially localized in real space with a characteristic
length about $\lambda\sim\hbar v_F/\Delta_E$, where $v_F$ is
the Fermi velocity and $\Delta_E$ is the magnitude of the energy
gap.~\cite{Prodan} For the parameter set used in Fig.\ 3, $\lambda$
is estimated to be between $1$ to $2$ lattice constants. 
When the confining potential $V_i$ is varying
relatively slowly in space, i.e., $\xi\gg\lambda$, one can find that
$P\sigma_zP$ roughly commutes with the confining potential. 
In this case, the confining potential is of no
influence on the spectrum of $P\sigma_zP$. Since the spin spectrum 
has a gap in the bulk~\cite{sheng}, this gap remains to open on the
smooth edges, as seen from Fig.\ 3d. As a result, the energy gap has to close due to the
topological requirement, resulting in gapless edge modes, as
observed in Fig.\ 3b.

\begin{figure}[hbtp!]
\includegraphics[width=3.4in]{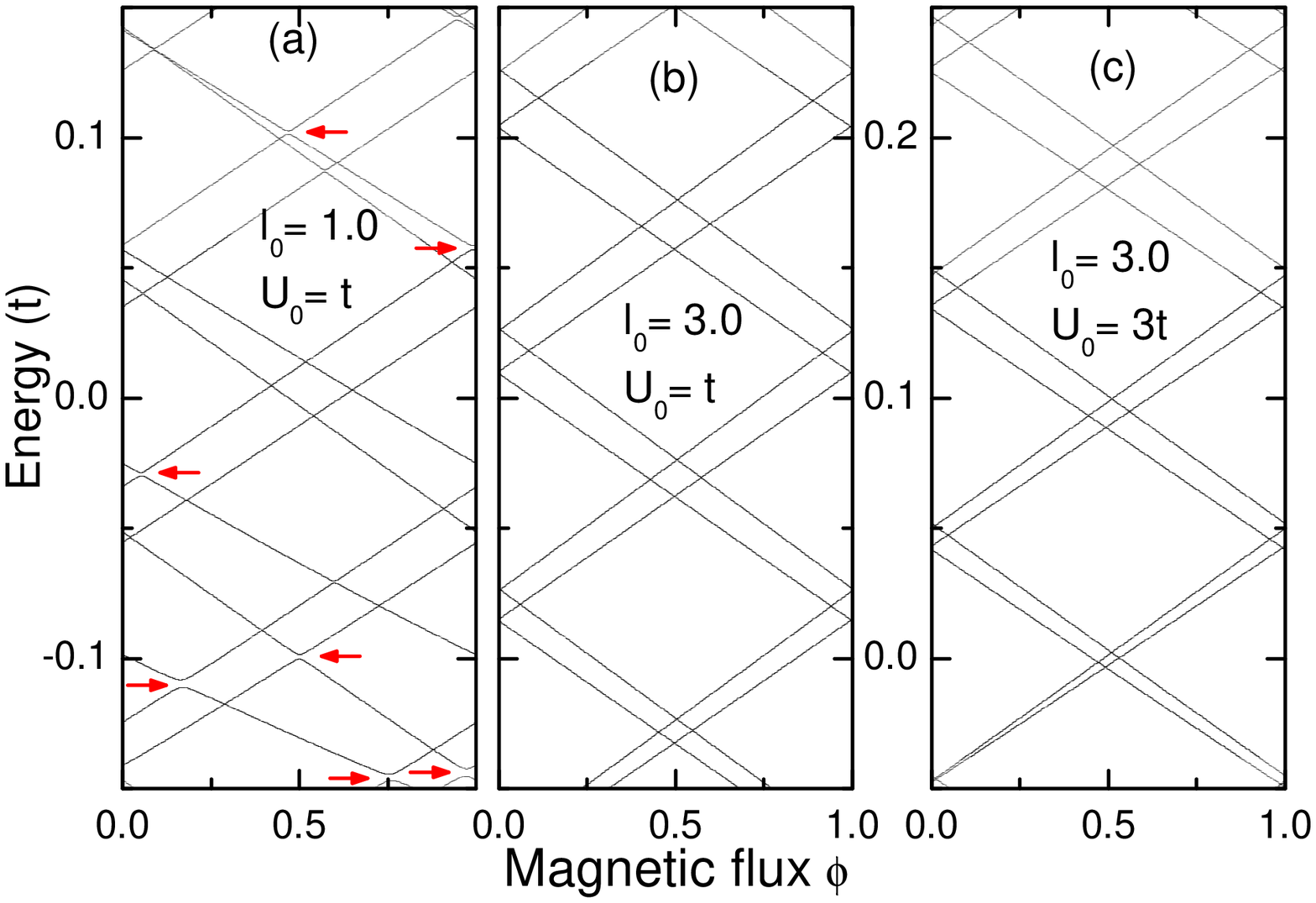}
\caption{(color online) Eigenenergies of the edge states as a function
of the magnetic flux $\phi$ 
threading the looped geometry with size
$120\times 60$ for (a) $l_0=1.0$, $U_0=t$, (b)
$l_0=3.0$, $U_0=t$, and (c) $l_0=3.0$, $U_0=3t$,
where the impurity concentration is fixed at $1\%$.
The other parameters are taken to be $V_{SO}=V_{R}=g=0.1t$,
$V_0=12t$, and $\xi=4$. 
Arrows in (a) indicate some of the relatively large energy gaps.}
  \label{Fig:disorder}
\end{figure}
Finally, we wish to discuss the robustness of the gapless
edge states found in the present TRS-broken QSH system. 
We consider a $N_x\times N_y$ sample forming 
a looped geometry as that
shown in Fig.\ 1.  $N_{I}$ 
nonmagnetic impurities are assumed to be 
randomly distributed in the sample at positions 
${\bf R}_{\alpha}$ with $\alpha=1,\cdots N_{I}$.
An extra term $H_{I}=\sum_{i}w_ic_{i}^\dagger
c_{i}$ is added to the total Hamiltonian $H$
to describe the effect of the impurity scattering, 
where $w_i=\sum_\alpha U({\bf r}_i-
{\bf R}_\alpha)$ with ${\bf r}_i$ as the 
position of the $i$-th
atom site. The impurity scattering potential is taken
to be $U({\bf r}_i-
{\bf R}_\alpha)=(U_0/l_0^2)\exp(-\vert {\bf r}_i-
{\bf R}_\alpha\vert/l_0)$ with $l_0$ as the correlation length
and $U_0$ the strength of the scattering potential. 
By inclusion of $1/l_0^2$ in the prefactor, the area integral
of the impurity potential is set to be independent of $l_0$. Figure 4 shows the
evolution of the calculated eigenenergies of a $120\times 60$ system
in the band gap upon adiabatic insertion of a magnetic flux $\phi$ into the ring,
for three different impurity scattering potentials. 
The number concentration of the impurities is fixed at $1\%$. 
For a very short correlation length $l_0=1$, for which
the impurity potential is nearly uncorrelated from one site to another,
we see from Fig.\ 4a that at $U_0=t$,
the energy levels of the edge states avoid to cross each other
as they move close, resulting in
small energy gaps in the spectrum, as indicated by the arrows. 
This level repulsion behavior is a 
signature of the onset of backward scattering~\cite{spinchern}. When the
characteristic length $l_0$ is increased to $l_0=3$ with
$U_0=t$ fixed, corresponding
to a relatively smooth impurity scattering potential, 
all the energy gaps vanish,
as shown in Fig.\ 4b. The energy levels move in straight lines and continue
to cross each other, a clear indication of quenching of the backward scattering~\cite{spinchern}. 
Such a level crossing feature 
is intact when $U_0$ is increased up to $3t$ for fixed $l_0=3$, 
as shown in Fig.\ 4c.
We thus conclude that the edge states
remain to be robust in the presence of 
relatively smooth impurity scattering potential of 
intermediate strength. This result can be understood
based upon an argument similar to that in the pure case. When $l_0$
is greater than the characteristic length $\lambda$ of the projected spin
operator $P\sigma_zP$, the impurity scattering potential  
nearly commutes with $P\sigma_zP$, and hence does not
affect much the spin spectrum gap, so that the energy gap
needs to close on the edges, which explains the level crossing behavior
of the edge modes.

In summary, based upon a general topological argument without
relying on the TRS or other symmetries,
we show that in a QSH system either the energy gap
or the gap in the spectrum of
$P\sigma_zP$ needs to close on the edges. 
We find that a TRS-broken QSH system can have either gapless or gapped
edge states, depending on the properties of the confining potential near
the boundaries. The gapless edge states are protected
by the bulk topological invariant rather than any symmetries,
which can remain to be robust in the presence of impurities.

\textbf{Acknowledgment} This work is supported by the State Key
Program for Basic Researches of China under Grant Nos. 2009CB929504
(LS), 2011CB922103 and 2010CB923400 (DYX), by the National Natural
Science Foundation of China under Grant Nos. 11074110 (LS), 11174125,
11074109, and 91021003 (DYX), and by a project funded by the PAPD of
Jiangsu Higher Education Institutions.

\end{document}